\documentclass[10pt,twocolumn]{article} 
\usepackage{graphicx}
\usepackage{url,hyperref}
\usepackage{amsmath}
\usepackage{amssymb}
\usepackage{amsfonts}
\usepackage{algorithmic}
\usepackage{textcomp}
\usepackage{xcolor}
\usepackage{siunitx}
\usepackage{todonotes}
\usepackage{float}
\usepackage{caption}
\usepackage{subcaption}
\usepackage{soul}
\usepackage{array}  
\usepackage{booktabs} 
\usepackage{titlesec}

\font\tenhv  = phvb at 10pt

\def\@maketitle
   {
   \newpage
   \null
   \vskip .375in 
   \begin{center}
      {\Large \bf \@title \par} 
      \vspace*{24pt} 
      {
      \large 
      \lineskip .5em
      \begin{tabular}[t]{c}
         \@author 
      \end{tabular}
      \par
      } 
      \vskip .5em 
      {
       \large 
      \begin{tabular}[t]{c}
         \@affiliation 
      \end{tabular}
      \par 
      \ifx \@empty \@email
      \else
         \begin{tabular}{r@{~}l}
            E-mail: & {\tt \@email}
         \end{tabular}
         \par
      \fi
      }
      \vspace*{12pt} 
   \end{center}
   } 

\def\abstract
   {%
   \centerline{\large\bf Abstract}%
   \vspace*{12pt}%
   }

\def\affiliation#1{\gdef\@affiliation{#1}} \gdef\@affiliation{}

\def\email#1{\gdef\@email{#1}}
\gdef\@email{}

\long\def\@makecaption#1#2{
   \vskip 10pt
   \setbox\@tempboxa\hbox{\tenhv\noindent #1.~#2} 
   \setlength{\@ctmp}{\hsize}
   \addtolength{\@ctmp}{-\@figindent}\addtolength{\@ctmp}{-\@figindent} 
   \ifdim \wd\@tempboxa >\@ctmp
      \begin{list}{}{\leftmargin\@figindent \rightmargin\leftmargin} 
         \item[]\tenhv #1.~#2\par
      \end{list}
   \else
      \hbox to\hsize{\hfil\box\@tempboxa\hfil} 
   \fi}

\begin{document}

\title{Random Telegraph Noise of a 28-nm Cryogenic MOSFET in the Coulomb Blockade Regime}

\author{HeeBong Yang$^{1*}$, Marcel Robitaille$^{1*}$, Xuesong Chen$^{1*}$, Hazem Elgabra$^{1}$,\\
Lan Wei$^{1}$, and Na Young Kim$^{1}$\\
    \thanks{$^*$ The authors equally contribute to the work.}
    $^1$Institute for Quantum Computing, Department of Electrical and Computer Engineering,\\ Waterloo Institute for Nanotechnology, University of Waterloo,\\ Waterloo, ON N2L 3G1, Canada
}
\date{}

\maketitle
\thispagestyle{empty}

\begin{abstract}
We observe rich phenomena of two-level random telegraph noise (RTN) from a commercial bulk 28-nm p-MOSFET (PMOS) near threshold at 14 K, where a Coulomb blockade (CB) hump arises from a quantum dot (QD) formed in the channel. Minimum RTN is observed at the CB hump where the high-current RTN level dramatically switches to the low-current level. The gate-voltage dependence of the RTN amplitude and power spectral density match well with the transconductance from the DC transfer curve in the CB hump region. Our work unequivocally captures these QD transport signatures in both current and noise, revealing quantum confinement effects in commercial short-channel PMOS even at 14 K, over 100 times higher than the typical dilution refrigerator temperatures of QD experiments ($<$100 mK). We envision that our reported RTN characteristics rooted from the QD and a defect trap would be more prominent for smaller technology nodes, where the quantum effect should be carefully examined in cryogenic CMOS circuit designs.
\end{abstract}

\section{Introduction}
Random telegraph noise (RTN) refers to two-level single-charge signal fluctuations and has been intensively studied for many decades in metal–oxide–semiconductor field-effect transistors (MOSFETs)~\cite{Machlup1954, Ralls1984, Kandiah1989, Kirton1989, Wang2018, Martin-Martinez2020}. The root cause of RTN in most MOSFETs is the capture and emission of a charged particle (i.e. an electron or a hole) by a defect (e.g. a charge trap) at the oxide layer interface or within the oxide layer. State-of-the-art commercial technology has further reduced the MOSFET size into deep sub-SI{100} regime. This transistor scaling exacerbates the charge trap phenomenon which significantly affects the device operation and performance. Recently, there are growing demands to develop large-scale fault-tolerant quantum computers based on short-channel complementary metal-oxide-semiconductor (CMOS) devices operating at cryogenic temperatures~\cite{Veldhorst2017}. Two physical parameters of CMOS, sub-micron size and low operation temperatures, trigger the observation of quantum  transport phenomena such as the Coulomb blockade (CB) and resonant tunneling appearing in small-size ($<$\SI{100}{\nano\meter}) quantum dots (QDs)~\cite{Averin1986, Fulton1987}. Furthermore, it is crucial to investigate the electrical transport noise properties of commercial MOSFETs in RTN, along with the CB oscillation, which was already reported in a silicon QD~\cite{Peters1999}, CMOS~\cite{Li2017, Li2018}, and individual carbon nanotubes~\cite{Jhang2014} below \SI{4}{\kelvin}. Recently, 22-nm and 40-nm CMOS devices are characterized at low temperatures, 2 K and 50 mK to exhibit QD transport signatures in Refs. \cite{Bonen2018} and~\cite{Yang2020}, respectively. However, noise measurements were not done in these sub-micron commercial CMOS at low temperatures yet. Here we report the RTN responses in the CB regime with commercial 28-nm MOSFETs at \SI{14}{\kelvin}. This work deepens the understanding of commercial CMOS noise and their uses for cryogenic applications. 

\section{Energy Band Diagram and Coulomb Blockade}
Among several foundry 28-nm PMOS and NMOS devices, we present our experimental data with a PMOS transistor with gate width $W$ = 1.2 $\mu$m and gate length $L$ = 28 nm as the device under test (DUT). The device sits at \SI{14}{\kelvin} in a dry cryostat (Advanced Research Systems, Inc). The average current-voltage ($I_{\text{D}}-V$) characteristics are measured along with the temporal drain-source current $I_{\text{D}}$ fluctuations at given gate-source voltage $V_{\text{GS}}$ and drain-source voltage $V_{\text{DS}}$ using a parameter analyzer (HP4156).

\begin{figure}[!h]
    \centering
    \includegraphics[width=1\linewidth]{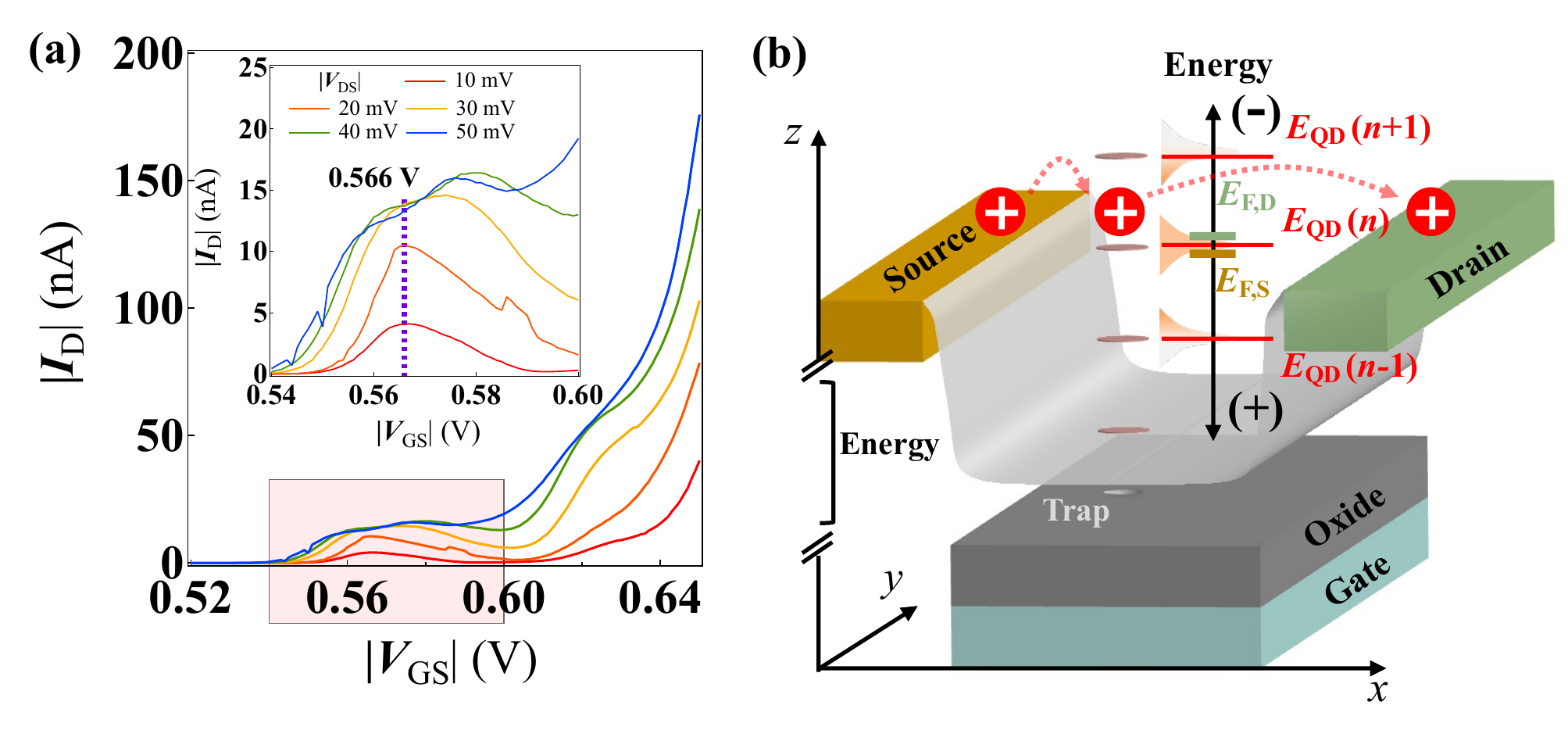}
    \caption{(a) The DC transfer curves $ |I_{\text{D}}|-|V_\text{GS}|$ of a 28-nm PMOS at $T$ = \SI{14}{\kelvin}. The inset displays zoomed-in view between 0.54 and 0.60 V showing a hump due to a QD. (b) The energy level configuration of the PMOS and the QD separated by tunnel barriers. Holes tunnel from source to drain via the resonant QD state $E_{\text{QD}}(n)$ (dish-like).} 
    \label{fig1:schematic_device}
\end{figure}

Fig.~\ref{fig1:schematic_device}(a) plots the DC transfer curves ($|I_{\text{D}}| - |V_\text{GS}|$) for five different $|V_{\text{DS}}|$ values at \SI{14}{\kelvin}. Interestingly, a hump is observed in the vicinity of the near threshold voltage ($|V_\text{GS}|$ = \SI{0.566}{\volt}) as shown in the inset of Fig.~\ref{fig1:schematic_device}(a). This hump, however, is more prominent at low drain voltages ($|V_{\text{DS}|} \leq$ \SI{50}{\milli\volt}), analogous to the signatures observed in \SI{22}{\nano\meter} CMOS at \SI{2}{\kelvin}~\cite{Bonen2018}, commercial bulk \SI{40}{\nano\meter} MOSFETs at \SI{50}{\milli\kelvin} \cite{Yang2020}, and in \SI{55}{\nano\meter} and \SI{75}{\nano\meter} CMOS devices at 2-5 K~\cite{Li2017, Li2018}. 

The hump shapes in the transfer characteristics (Fig.~\ref{fig1:schematic_device}(a)) can be well attributed to the formation of QDs in the channel. In the weak inversion region, the space charges exist close to the source (S) and drain (D), and the slightly inverted channel forms a QD with quantized energy states ($E_{\text{QD}}(n)$ in Fig.~\ref{fig1:schematic_device}(b)). At $|V_\text{GS}|$ = \SI{0.566}{\volt}, current is at the local maximum due to resonant tunneling when $E_{\text{QD}}(n)$ aligns with the Fermi energy at S and D terminals ($E_\text{F,S}$ and $E_\text{F,D}$ in Fig.~\ref{fig1:schematic_device}(b)), while no current flows if no QD state resides within the $|V_\text{DS}|$ bias window. As a consequence, the CB hump appears in the inset of Fig.~\ref{fig1:schematic_device}(a). Here, a notable current hump is attributed to the larger $|V_\text{DS}|\ge 10$ mV and higher temperature \SI{14}{\kelvin} compared to typical QD experiments with 1-5 mV at 1-2 K in~\cite{Peters1999}. A current wiggle around $|V_\text{GS}| \sim \SI{0.625}{\volt}$ would be the next CB hump; however, it is not prominent due to more carriers flowing in the channel. Eventually, the QD-associated humps disappear in the saturation regime.

Following similar procedures of the standard CB data analysis as described in~\cite{Yang2020}, we can extract series of capacitance ($C$) to three terminals of the PMOS: gate-capacitance $C_{\text{G}}$ = \SI{2.28}{\atto\farad} from the gate voltage separation between adjacent CB humps of 70 mV, source-capacitance $C_{\text{S}}$ = \SI{2.75}{\atto\farad}, and drain-capacitance $C_{\text{D}}$ = \SI{1.37}{\atto\farad} from two slopes of the DC curves in Fig.~\ref{fig1:schematic_device}(a). The location of the QD in the \SI{28}{\nano\meter} channel is about \SI{9}{\nano\meter} from S, which is estimated by the ratio between $C_{\text{S}}$ and $C_{\text{D}}$. We assume that the QD is a disk shaped, and with the oxide thickness of \SI{1.3}{\nano\meter} and SiO$_2$ relative permittivity, the size of the QD is also theoretically calculated to be a radius as about \SI{5.96}{\nano\meter} which is sketched in Fig.~\ref{fig1:schematic_device}(b).

\begin{figure}[!h]
    \centering
    \includegraphics[width=1\linewidth]{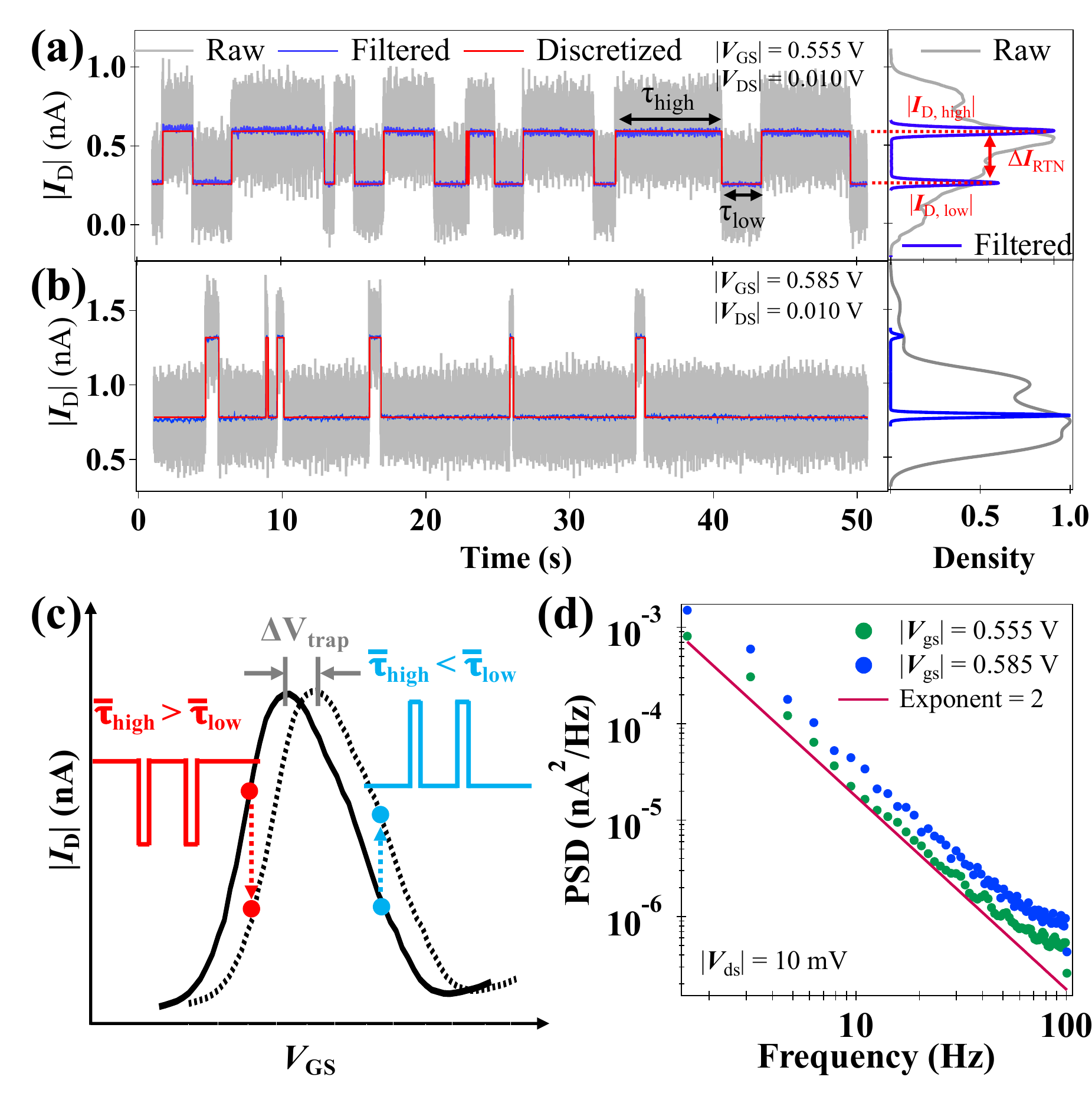}
    \caption{Sampling and histogram of $I(t)$ at $|V_\text{DS}|$ = \SI{10}{\milli\volt} (a) $|V_\text{GS}|$ = \SI{0.555}{\volt} and (b) $|V_\text{GS}|$ = \SI{0.585}{\volt}: Raw data (gray), filtered (rolling-averaged) data (blue), and discretized data (red). (c) The trapping behaviors are illustrated with empty trap (solid line) and occupied trap (dashed line) across the hump area of the $|I_\text{D}|$-$|V_\text{GS}|$ sweep.
    (d) The current power spectral density (PSD) of the discretized data in Figs. 2(a) and 2(b) with the $1/f^2$ PSD guideline in red.}
\label{fig2:sampling}
\end{figure}

Temporal current $|I_{\text{D}}(t)|$ traces are recorded at the sampling interval of \SI{4.96}{\milli\second} (the minimum time interval in HP4156 for the current value range (0.5-14 nA) without any artificial digitization) in the CB hump region (\SI{0.54}{\volt} $<|V_{\text{GS}}| <$ \SI{0.60}{\volt}), and typical $|I_{\text{D}}(t)|$ data are presented at the left and right sides of the maximum hump at $|V_{\text{GS}}| =$ \SI{0.566}{\volt} in Figs.~\ref{fig2:sampling}(a) and ~\ref{fig2:sampling}(b), respectively. Random jumps between two unique current levels ($|I_\text{D, high}|$ and $|I_\text{D, low}|$) on top of background white-noise fluctuations occur distinctively and persistently in the whole time window even in the raw data (gray in Figs.~\ref{fig2:sampling}(a) and (b)). To remove featureless background noise, we apply a rolling average to obtain the filtered blue signals. Consequently, the kernel density estimation method~\cite{Rosenblatt1956} applied to the filtered signal creates discretized RTN signals in red, which capture two sharp occurrence humps separated by the RTN amplitudes $\Delta I_\text{RTN}$. Two time constants, $\tau_\text{high}$ and $\tau_\text{low}$, are defined as the dwell time at $|I_\text{D, high}|$ and $|I_\text{D, low}|$, respectively. The observed transient RTN in the temporal domain can be attributed to a trap at the oxide-channel interface or within the oxide, which captures a hole from the PMOS channel or emits a hole into the PMOS channel. The trapping behavior essentially causes a horizontal shift of the DC transfer curve by a constant amount ($\Delta V_{\text{trap}}$) between the trap is empty (solid line) and occupied (dashed line) (Fig.~\ref{fig2:sampling}(c)). We examine the current noise power spectral density (PSD) that exhibits the  $1/f^2$-Lorentzian signature from a single RTN~\cite{Kogan1996} in Fig.~\ref{fig2:sampling}(d).

\section{RTN Behavior around CB hump}

\begin{figure}[!t]
    \centering
    \includegraphics[width=1\linewidth]{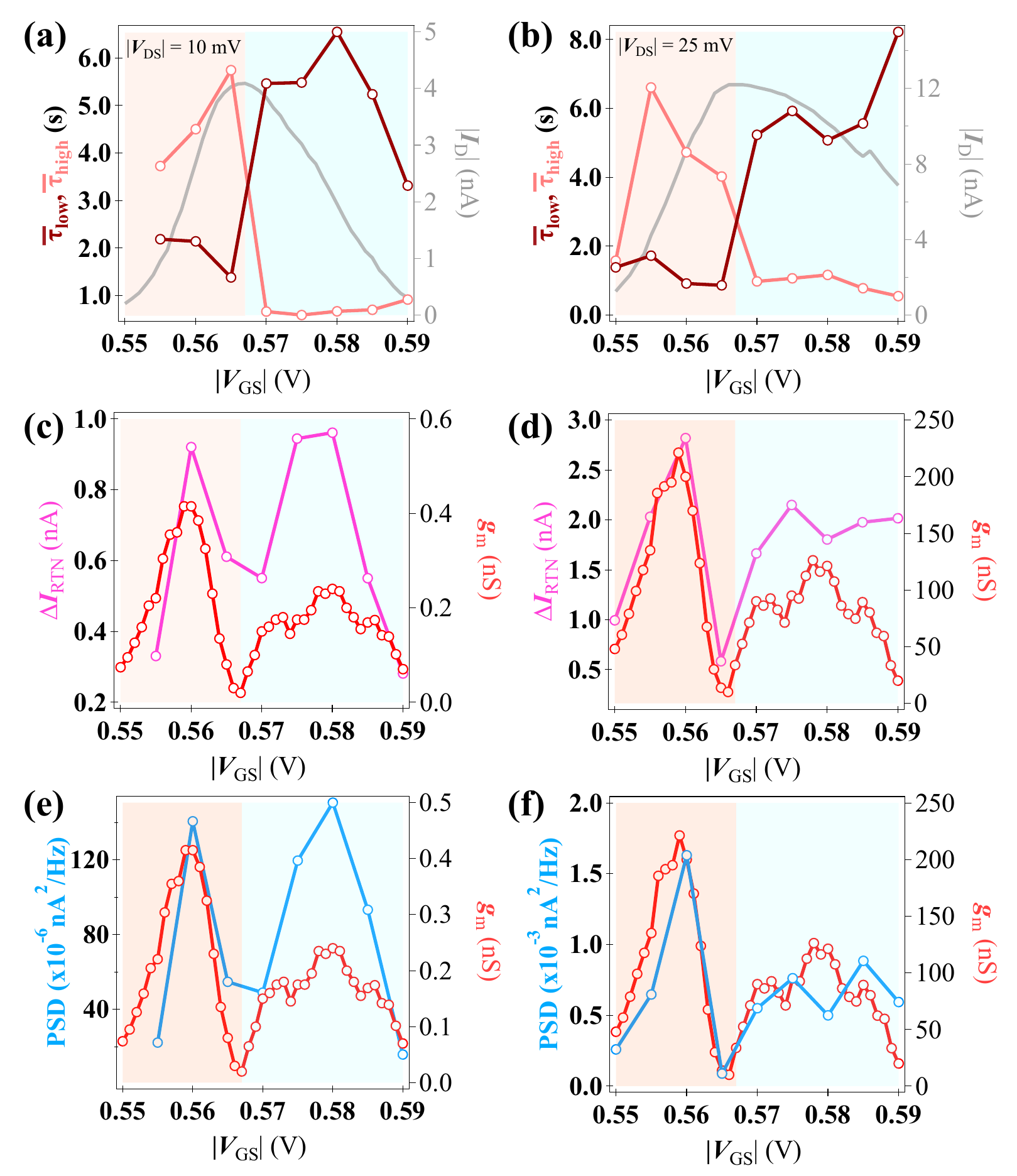}
    \caption{$|V_{\text{GS}}|$ dependence of RTN time constant $\overline{\tau}_{\text{high}}$, $\overline{\tau}_{\text{low}}$, $\Delta I_{\text{RTN}}$, and RTN PSD at \SI{10}{\hertz} at $|V_{\text{DS}}| = \SI{10}{\milli\volt}$ (a, c, e) and $|V_{\text{DS}}| = \SI{25}{\milli\volt}$ (b, d, f). Note that the gray $|I_\text{D}|$ in (b) is averaged data from $|V_{\text{DS}}|$ = $\SI{20}{\milli\volt}$ and $\SI{30}{\milli\volt}$.  (c) and (d) are $\Delta I_\text{RTN}$ (pink, left $y$-axis) and $g _{\text{m}}$ (red, right $y$-axis), and (e) and (f) are PSD and $g_\text{m}$.}
\label{fig3:Vg_depen}
\end{figure}

The average $\bar{\tau}_\text{high}$ and $\bar{\tau}_\text{low}$ from discretized signals exhibit dramatic switching from $|I_\text{D, high}|$-dominance to $|I_\text{D, low}|$-dominance around $|V_\text{GS}|$ = 0.566 V. This voltage corresponds to the humps of the DC $I_\text{D}-V_\text{GS}$ variations (gray solid line) in Figs.~\ref{fig3:Vg_depen}(a) and (b), where the QD energy levels align with $E_{\text{F,S}}$ or $E_{\text{F,D}}$. Below $|V_\text{GS}|$ = 0.566 V (light orange region), the device prefers the $|I_\text{D, high}|$ level with longer $\bar{\tau}_\text{high}$. For $|V_{\text{GS}}| > $ \SI{0.566}{\volt}, a drastic switching of the time constants occurs in $|V_{\text{DS}}|$ = \SI{10}{\milli\volt} and \SI{25}{\milli\volt} (Figs.~\ref{fig3:Vg_depen}(a) and (b)). 

$\Delta I_\text{RTN}$ is displayed against $|V_\text{GS}|$ in Figs.~\ref{fig3:Vg_depen}(c) and (d). $\Delta I_\text{RTN}$ oscillates at half period of the CB hump broadening in $|V_\text{GS}|$, where the sharp minimum coincides with $|V_\text{GS}| \sim \SI{0.566}{\volt}$. The unclear second hump of $\Delta I_\text{RTN}$ in Fig.~\ref{fig3:Vg_depen}(d) may be due to the larger bias voltages  with more charges in the channel to screen the quantum behavior. This $\Delta I_\text{RTN}-|V_\text{GS}|$-oscillation is exactly the same as the RTN amplitude behavior in a silicon QD reported in~\cite{Peters1999}, where the QD surface potential $\phi$ is controlled by $|V_\text{GS}|$ but disturbed by a trap nearby. In addition, for the DUT biased near threshold region, $\phi$ is linearly modulated by $|V_\text{GS}|$. We could express $\Delta I_\text{RTN}$ as the uniform potential fluctuation model,
\begin{equation}
	\Delta I_\text{RTN} = \frac{\partial I_\text{RTN}}{\partial \phi} \Delta \phi = \frac{\partial I_\text{D}}{\partial V_\text{GS}} \Delta V_\text{GS} = g_\text{m} \Delta V_\text{GS},
	\label{eq:delta_i}
\end{equation}
where $g_\text{m}$ is the transconductance and $\Delta V_\text{GS}$ is the equivalent fluctuation with respect to $V_\text{GS}$ caused by the capture and emission behavior of the trap. Under this picture, the capture and emission behavior causes a constant shift to the DC transfer curve (Fig.~\ref{fig2:sampling}(c)), whereas the $g_\text{m}$ directly modulates its effect on $\Delta I_\text{RTN}$. The theory can explain the dramatic switching of the time constant: the $g_\text{m}$ changes its sign from positive to negative at the CB hump. Furthermore, $\Delta I_\text{RTN}$ and $g_\text{m}$ are overlapped along $|V_\text{GS}|$ (Figs.~\ref{fig3:Vg_depen}(e) and (f)). Across the CB hump, the size of $\Delta I_\text{RTN}$ increases 4-5 times bigger from the minimum value, which reflects the large $g_\text{m}$ change. This is unseen near threshold in classical operations.

One prominent observation is the high $\Delta I_\text{RTN}/|I_\text{D}|$, over 50\% of $|I_\text{D}|$, which implies the significant RTN impact the charge transport, in particular when the number of charges $N$ in the channel is small.  $N$ is estimated theoretically by the inverse of $\Delta I_\text{RTN}/|I_\text{D}|$  described in~\cite{Hung1990}. $N$ is around 1-5 at $|V_\text{GS}| = 0.56 $ V, but it grows quickly to 20-40 at $|V_\text{GS}| = 0.57 $ V, consistent with $N \sim $ 10-20 estimated from the thermal energy hopping picture. The charged trap may generate an additional electric-field to push the transfer curve, which demonstrates that the device acts like a quantum single-electron transistor to sensitively detect environmental conditions. Hence, quantum transport properties should be taken into account when designing cryogenic CMOS circuits, especially for devices that are biased at low voltage.

Figs.~\ref{fig3:Vg_depen}(e) and (f) give the PSD values at 10 Hz as a function of $|V_\text{GS}|$. The half-period PSD oscillations in both $|V_\text{DS}|$ nicely match with the $g_m$-$|V_\text{GS}|$ trend, which further supports the previous uniform potential fluctuation model. The interesting observation of the minimum noise at the maximum of $I_\text{D}$ in the hump region offers an optimal bias condition with minimum RTN to operate cryogenic devices.

\section{Conclusion}

We observe quantum footprints in the RTN behavior of foundry 28-nm bulk PMOS devices even at 14 K arising from CB and resonant tunneling through systematic analysis of fluctuating currents in both time and frequency domains to quantify the RTN parameters. A uniform potential fluctuation model is proposed to explain the dramatic RTN switching behavior around the CB hump. Our complete DC and noise studies of commercial MOSFETs support potential cryo-applications including CMOS quantum computers.

\section*{Acknowledgment}

We thank Industry Canada, the Ontario Ministry of Research \& Innovation through Early Researcher Awards (RE09-068, ER18-14-276) and the Canada First Research Excellence Fund-Transformative Quantum Technologies (CFREF-TQT). J. Watt and C. Chen in Intel for samples, A. Malcolm for early work and J. Baugh for helpful discussions are appreciated.

\bibliographystyle{ieeetr}
\bibliography{refs}
\end{document}